\documentclass[12pt]{iopart}

\usepackage{cite}
\usepackage{graphicx}%

\begin{document}

\title[Soft-mask fabrication of GaAs nanomembranes]{Soft-mask fabrication of gallium arsenide nanomembranes for integrated quantum photonics} 

\author{L Midolo, T Pregnolato, G Kir\v{s}ansk\.{e} and S Stobbe}
\address{Niels Bohr Institute, University of Copenhagen, Blegdamsvej 17, DK-2100 Copenhagen, Denmark}
\ead{midolo@nbi.ku.dk}

\begin{abstract}
We report on the fabrication of quantum photonic integrated circuits based on suspended GaAs membranes. The fabrication process consists of a single lithographic step followed by inductively-coupled-plasma dry etching through an electron-beam-resist mask and wet etching of a sacrificial layer. This method does not require depositing, etching, and stripping a hard mask, greatly reducing fabrication time and costs, while at the same time yielding devices of excellent structural quality. We discuss in detail the procedures for cleaning the resist residues caused by the plasma etching and present a statistical analysis of the etched feature size after each fabrication step.  
\end{abstract}



\maketitle 

\section{Introduction}
Quantum photonic integrated technology deals with the generation, manipulation and detection of quantum light, typically in the form of single photons, within a semiconductor chip \cite{shields_semiconductor_2007,obrien_photonic_2009,rogers_synthesis_2011,benson_assembly_2011,lodahl_interfacing_2013,metcalf_quantum_2014,silverstone_-chip_2014}. It constitutes an active field of research aiming at solving scalability issues of free-space quantum-optics experiments by means of optical fields confined in dielectric waveguides and resonant nanostructures such as photonic crystals (PhC). In particular, the efficient and coherent generation of single photons within direct-bandgap semiconductors constitutes a major step forward in realizing a photonic platform for quantum computation. To this end, InAs quantum dots (QDs) in GaAs, embedded in photonic crystals \cite{shields_semiconductor_2007,lodahl_interfacing_2013}, have proven to be an excellent light-matter interface for near-unity efficiency single-photon generation \cite{hoang_enhanced_2012, arcari_near-unity_2014}, nonlinear interactions \cite{fushman_controlled_2008, javadi_single-photon_2015}, and cavity quantum electrodynamics \cite{hennessy_quantum_2007}. To achieve the high efficiencies needed for generating and routing photons, photonic integrated circuits have to be fabricated with highly precise micro- and nano-fabrication techniques. These techniques involve the direct writing of nanostructures using electron-beam lithography into a resist mask and the subsequent transfer to the substrate via anisotropic dry etching. The optical quality of the devices depends crucially on this last step, which is usually carried out using reactive-ion etching (RIE).

\begin{figure}
\centering
\includegraphics{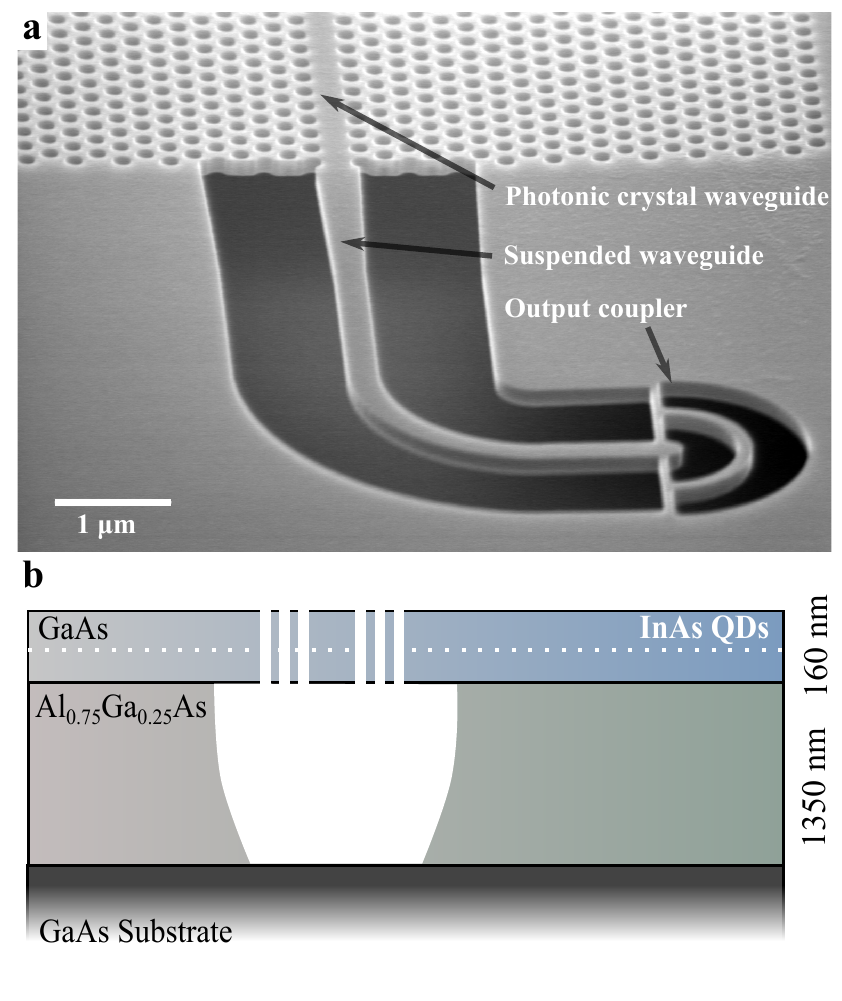}
\caption{\label{fig1}Integrated quantum photonic circuits in GaAs membranes. (a) Scanning electron micrograph of a PhC waveguide coupled to a suspended waveguide, fabricated in a GaAs nanomembrane. (b) Layout of the sample, showing schematically how the membranes have been patterned by dry etching and selective wet etching.}
\end{figure}

In RIE processes, the choice of the mask material is of utmost importance in order to etch features deeply into the substrate while keeping a high structural quality. In particular, the mask largely determines the selectivity of the process, i.e., the ratio between the etch rates of the mask and the semiconductor. For silicon photonic circuits the electron-beam resist is deposited directly on the substrate and used as a mask for dry etching, yielding good selectivity \cite{loncar_design_2000}. For III/V semiconductor etching, and in particular for PhC fabrication, it is more common to use intermediate layers, so-called hard masks, such as Si$_3$N$_4$, SiO$_2$, or metals, to improve the selectivity of small features\cite{painter_two-dimensional_1999,gonzalez_fabrication_2014}. There are situations, however, where introducing an additional mask layer is not desirable, especially when oxides or metals have been already deposited on the sample for other purposes, e.g., for the fabrication of superconducting single photon detectors \cite{sprengers_waveguide_2011}. In this letter we present a soft-mask process where a circuit is etched directly into a GaAs membrane through the patterned electron-beam resist \cite{hennessy_positioning_2004,khankhoje_modelling_2010}. Unlike hard-mask processes, this method involves fewer fabrication steps, reduces the need for expensive deposition equipment and operates at lower temperatures. We have developed a fabrication recipe that allows realizing circuits as shown in figure\ \ref{fig1}(a), composed of PhC waveguides, suspended waveguides and other passive components such as power splitters and out-couplers. Suspended PhC waveguides are needed to achieve a high coupling efficiency between the emitter and the optical mode. Since layers with low refracive indices cannot be grown epitaxially on GaAs, there is a significant mode mismatch between light propagating in PhCs and in GaAs ridge waveguides. Therefore the entire circuitry is fabricated on suspended membranes. 

\section{Experimental procedure}
\subsection{Sample preparation}
The samples are grown by molecular-beam epitaxy on a (001) GaAs substrate with the material stack shown in figure\ \ref{fig1}(b). It consists of a 160-nm-thick GaAs membrane with a layer of self-assembled InAs QDs located in the center. The membrane is grown atop a 1350-nm sacrificial layer (Al$_{x}$Ga$_{1-x}$As, x=0.75). 
Since the QD emission is centered around 940 nm at 10 K, the sacrificial layer thickness is designed so that light exiting from the couplers, see figure \ref{fig1}(a), is emitted preferentially towards the top, while keeping a sufficient distance from the substrate to avoid breaking the symmetry of the optical confinement in the membrane. The aluminum fraction in the sacrificial layer is chosen so that it can be easily removed by hydrofluoric or hydrochloric acid. We have found incomplete etching when x$<$0.7 and excessive, uncontrollable undercut rates with x$>$0.8.
The sample is initially cleaned from dust and organic residues using, in sequence, acetone and isopropyl alcohol (IPA) and subsequently dehydrated at 185 $^\circ$C for 5 minutes. 
The electron-beam resist ZEP520A is deposited by spin coating at 2000 rotations per minute and baked at 185 $^\circ$C for 5 minutes on a hot plate to produce a mask thickness of 500--540 nm (measured by a spectral film-thickness analyzer). ZEP520A is chosen since it combines high resolution with good resistance to dry etching. 
The resist is patterned using a 100 kV electron-beam lithography tool (Elionix ELS-7000G) and developed in n-amylacetate for 1 minute at 22 $^\circ$C and rinsed in IPA for 10 s. PhC holes are exposed at a uniform dose of 300 $\mu$C/cm$^2$ whereas the remaining features are exposed at a lower dose of 230 $\mu$C/cm$^2$.

\subsection{Dry etching}
To etch the patterns vertically through the membrane, a RIE tool (Plasmalab 100) from Oxford Instruments, equipped with an inductively-coupled-plasma (ICP) source, is used. The system allows monitoring the remaining resist thickness in real-time using laser interferometry. We have observed that when the resist thickness in the homogeneous (un-exposed) part of the mask is reduced to less than 200 nm, the etched features tend to enlarge and erode around the sidewalls. This effect stems from the fact that, in the proximity of the exposed patterns, the resist etches faster, leaving the GaAs surface exposed to the plasma. Using the reflectance signal from the resist, it is possible to stop the process before this phenomenon occurs. 
Moreover, to avoid excessive heating and resist re-flow (occurring at around 140 $^\circ$C) due to the reactive plasma, it is important to establish a good thermal contact between the sample and the electrode. For this purpose, the sample is glued to a Si carrier wafer using thermally conducting and removable adhesives (fluorinated lubricants or wax) and the electrode temperature is lowered to 0 $^\circ$C. Helium flow is introduced to attain a good thermal contact between the carrier wafer and the electrode table throughout the entire process. 
A plasma chemistry based on BCl$_3$, Cl$_2$ and Ar is used.\cite{hennessy_positioning_2004} Chlorine is the main etching agent while the boron tetrachloride is used for side-wall passivation and argon for dilution \cite{atlasov_effect_2009, braive_inductively_2009}. Additionally, Ar can enhance the physical etching (sputtering) and reduce the sidewall roughness \cite{baca_fabrication_2005}.
The 65-mm-diameter ICP coil power is set to 300 W since higher values lead to lower selectivity. The electrode RF power is adjusted so that the voltage between the  RIE plate and the chamber walls is approximately around 300 V and that the plasma is stable. For most of the experiments, an RF power of 47 W is used. The chamber pressure is initially set to a strike pressure of 25 mTorr and then rapidly lowered to the desired value once the plasma has ignited. 
We have found that an optimum etching profile is obtained when the BCl$_3$/Cl$_2$/Ar flow ratio is set to 3/4/25 sccm, respectively and the chamber pressure is 4.7 mTorr. 

\subsection{Sacrificial-layer undercut}
After plasma etching, the residual resist layer is removed by soaking the sample for 10 minutes in n-methylpyrrolidone (NMP) at 60 $^{\circ}$C. Once cooled, NMP is removed by washing it in IPA. 
Two acids have been tested for the selective removal of the Al$_{0.75}$Ga$_{0.25}$As sacrificial layer, namely a 10\% solution of hydrofluoric acid (HF 40\% mixed 1:3 with water by volume) and a cold ($<$5 $^\circ$C) hydrocloric acid solution (HCl 37\% mixed 4:1 with H$_2$O) \cite{sun_selective_2009}. The HF etch rate is relatively high (approximately 65 nm/s lateral etch rate) whereas HCl provides a much lower rate ($\sim$3 nm/s depending on the solution temperature and Al content). 
To properly clean the sample from residues left by wet and dry etching, a 60 s hydrogen peroxide dip followed by deoxidation in a 20\% potassium hydroxide solution (25 g of KOH dissolved in 100 ml water) is used.  
Suspended GaAs structures consisting of thin membranes usually show low mechanical stiffness, which can lead to a collapse onto the substrate due to capillary forces and subsequent inter-solid adhesion \cite{maboudian_critical_1997}. To avoid such stiction failures, attention is paid throughout the entire wet etching procedure to keep the surface covered by a liquid droplet until the end of the process. Then the sample is dried either by immersion in IPA followed by evaporation or by a supercritical point dryer in carbon dioxide. 

\section{Results and discussion}

\subsection{Dry etching}
The samples are patterned with PhC waveguides (hole-to-hole distance 245 nm) coupled to suspended waveguides to simultaneously inspect the quality of different aspect-ratio features (defined as the average feature size over mask thickness). Plasma etching of small aspect-ratio features usually causes what is known as RIE depth lag, i.e. a correlation between the feature size and its etch depth. This implies that the optimum etching parameters depend on the feature size. Insufficient etch depth leads to incomplete penetration through the membrane and prevents the subsequent undercut of the membrane. 
\begin{figure}
\centering
\includegraphics{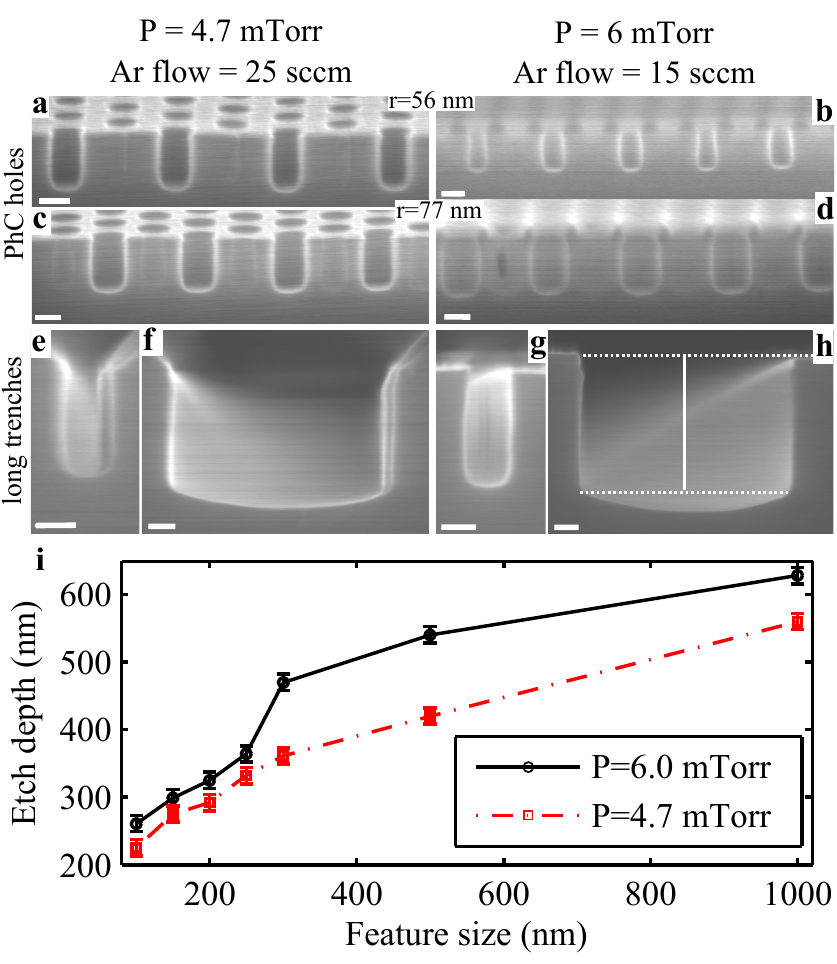}
\caption{\label{fig2} Optimization of dry-etching parameters. (a-h) Scanning electron micrographs of dry-etched GaAs membranes with different patterns. The process with low (high) pressure and high (low) Ar flow is shown to the left (right). (a-d) PhC holes with different radii and (e-h) cross-section of long trenches with different widths (100 nm and 1 $\mu$m). (i) Etching depth as a function of the trench size for two etching conditions. The depth is a function of the feature size are extracted as indicated in (h) by solid and dashed lines, respectively. All scale bars are 100 nm.}
\end{figure}
The ICP-RIE process is optimized starting from values in the literature \cite{hennessy_positioning_2004, braive_inductively_2009}. The parameter space has been  narrowed down by optimizing the BCl$_3$/Cl$_2$/Ar ratio first and then by studying the effect of the chamber pressure and Ar flow only. Two processes have been optimized, one dedicated to vertical etching of PhC, and one for deeper etching of large features. 
In figure\ \ref{fig2} these two recipes are compared in terms of etch depth and quality. Higher pressures (6 mTorr) lead to significantly higher selectivity, but more curved sidewall profiles. This is visible in particular when examining the cross-sectional profiles of PhC holes by scanning electron microscopy (SEM) as shown in figure\ \ref{fig2}(a-d). To achieve a better verticality the pressure is reduced to 4.7 mTorr while the Ar flow is increased to control the sidewall roughness and the selectivity. As the pressure is reduced, physical etching dominates over chemical etching, resulting in lower selectivity. With these settings the resist etch rate is on average around 280 nm/min while the total etching time is between 45 and 55 s depending on the initial resist thickness. The smallest holes that can be etched deeper than 160 nm (the membrane thickness) have a radius of 56 nm and a corresponding etch rate of approximately 190 nm/min, resulting in a selectivity (GaAs etch to mask) of $\sim$2:3. For larger features such as the trenches shown in figure \ref{fig2}(f) and \ref{fig2}(h), the selectivity grows beyond 2:1. Figure\ \ref{fig2}(i) summarizes the etch depth as a function of the width of long trenches whose cross-sections are shown in figures\ \ref{fig2}(e-h) for the two recipes. As expected, the 6 mTorr recipe produces deeper features and it is thus more suitable for samples containing large non-periodic features where a deeper etching is needed. For PhC etching, however, the low-pressure recipe yields better results in terms of sidewall smoothness and verticality. 

\subsection{Wet etching}
\begin{figure}\centering
\includegraphics{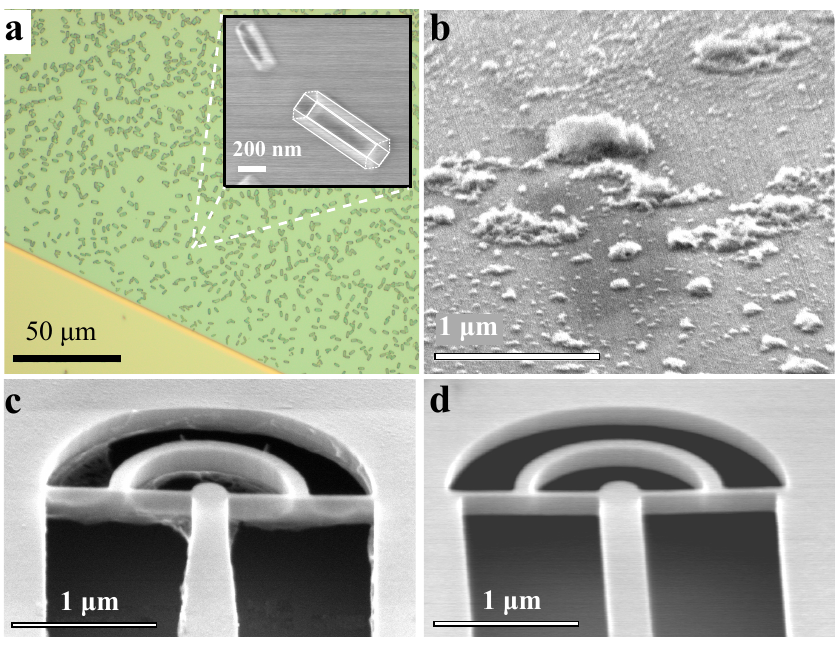}
\caption{\label{fig3}Removal of residues from etching leaving nanostructures of high structural quality. (a) Optical microscope image of AlF$_3$ crystals. Inset: SEM micrograph of the crystals where the crystalline structure has been outlined with solid white lines. SEM micrographs of (b) Al(OH)$_3$ residues on GaAs, and (c) debris of a thin, probably carbon-rich, film left by dry etching sticking to an output coupler. (d) Same as (c), after the treatment with H$_2$O$_2$ and KOH as described in the text.}
\end{figure}

To reduce losses due to scattering in waveguides, it is crucial to properly clean the devices after wet etching. Several residues are observed on the fabricated structures after their release, especially when HF is used to remove Al(Ga)As. 
Most of the byproducts of Al(Ga)As and HF are gaseous or soluble in water \cite{voncken_etching_2004} but we have noticed that solid compounds such as aluminum trifluoride (AlF$_3$), aluminum hydroxide (Al$($OH$)_3$), and other amorphous residues caused by dry etching may be deposited at the bottom of the substrate and around the etched features. AlF$_3$ is identified by its crystalline structure as shown in figure\ \ref{fig3}(a) and it can be easily dissolved by a 10 min rinse in water. Al(OH)$_3$ residues, see figure\ \ref{fig3}(b), are sometimes observed as a product of HF vapors and can be removed by a dip in 20\% potassium hydroxide.   
Figure\ \ref{fig3}(c) shows an amorphous residue that sticks to the edges of the released GaAs structures and which is always observed at the end of the wet etch process. Such fragments of what appears to be a thin organic layer cannot be removed by further HF etching. Moreover, it is also observed after HCl etching, indicating that it is more likely a product of dry etching. 
To investigate the origin of this compound and, more importantly, ways of removing it, different cleaning steps before and after the HF undercut are performed. 
Using basic or acidic de-oxidizing solutions (such as dilute phosphoric acid, dilute hydrocloric acid, potassium hydroxide or ammonium hydroxide) does not help in removing the substance. 
Instead, a single digital etching cycle \cite{desalvo_wet_1996}, which consists of 60 s immersion in concentrated hydrogen peroxide (H$_2$O$_2$ 30\%) followed by a 60 s water rinse and 120 s etching in 20\% potassium hydroxide, can remove the residues completely, leaving a smooth surface as shown in figure\ \ref{fig3}(d) and \ref{fig4}(a).
The use of a soft mask in very hot plasmas typically leads to resist damage and the formation of carbon-rich films, which are hard to strip afterward. These residues may re-deposit inside the etched features during the dry etching or later on, during the resist stripping in NMP. We speculate that this is the case for the residue in figure\ \ref{fig3}(c). This hypothesis is supported by the fact that applying an oxygen plasma seems to be beneficial, although not sufficient, to removing the debris of this carbon film. Additionally, we have noticed that using concentrated H$_2$O$_2$ alone, followed by a water rinse, removes the residue completely, suggesting that the carbon-rich film is oxidized by peroxide and turned into volatile compounds such as CO$_2$. 
Clearly, the use of peroxide has the side-effect of oxidizing GaAs and thus necessitates a final de-oxidation step. Among the various de-oxidizing agents available, KOH has the additional advantage of performing a thorough cleaning of other residues such as Al(OH)$_3$. 

\subsection{Effects of the process steps on the hole size}
\begin{figure}\centering
\includegraphics{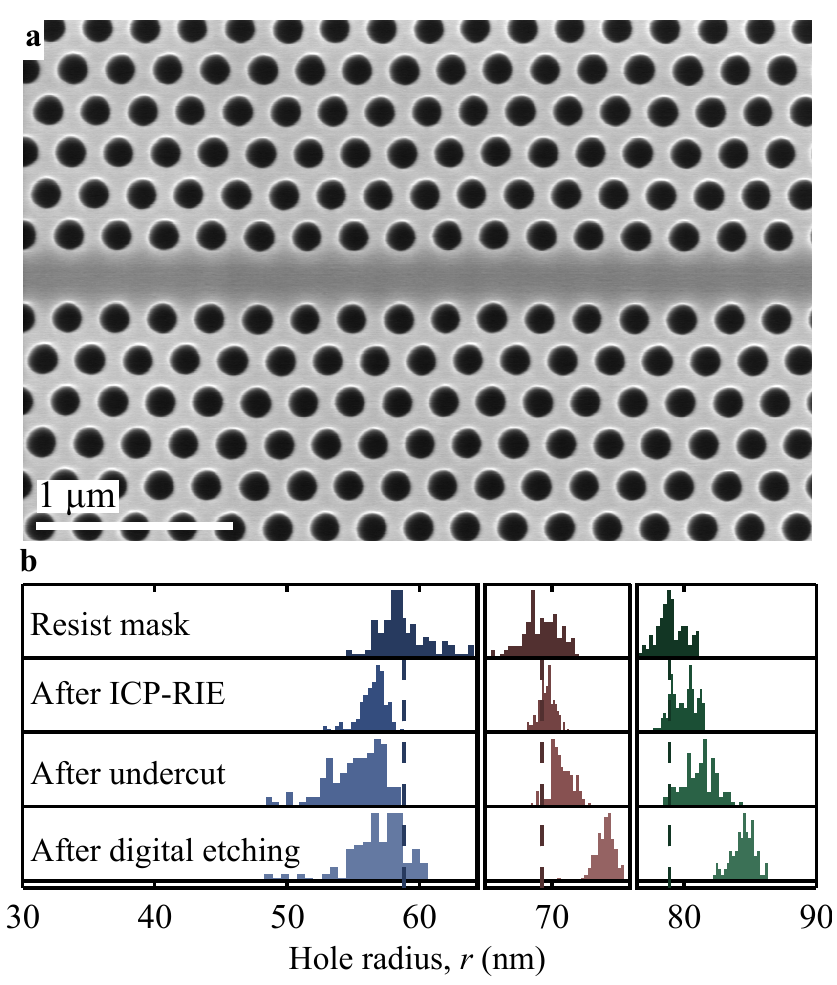}
\caption{\label{fig4}Reproducibility and accuracy of PhC patterning. (a) SEM micrograph of a PhC waveguide (PhC lattice parameter 245 nm and radius $r=70$ nm). (b) Statistics of the hole radius distribution before and after each etching step. The vertical dashed lines indicate the average radius value in the mask. The original mask design values are 61 nm (blue), 70 nm (red), and 77 nm (green).}
\end{figure}

The most delicate aspect in fabricating PhC waveguides is the accuracy in the hole positioning and radius, which is mainly achieved using state-of-the-art electron-beam-lithography tools. In particular, the accuracy in reproducing a given hole radius allows positioning the photonic band edge to the correct wavelength relative to the QD emission. To this end, the effects on the hole size due to processing are analyzed. Figure\ \ref{fig4}(b) summarizes the results of a statistical analysis of SEM micrographs of PhCs with different radii after each step. All images have been acquired with the same settings and equalized to uniform brightness and contrast. An enlargement of 2--3 nm is observed after the digital etching. This is in the range expected from the literature \cite{desalvo_wet_1996} and can easily be compensated by proper mask design. We note additionally that the dry-etching process does not visibly enlarge the hole radius. This indicates that the process is very accurate in transferring a pattern from a soft mask into GaAs. From figure\ \ref{fig4}(b) we also notice that when the holes in the mask are smaller than 60 nm, they tend to have a larger distribution, which partly obscures the enlargement from step to step. This higher radius spread is probably due to incomplete opening in the resist mask and has to be corrected by adjusting the exposure dose in the electron-beam lithography. 

\section{Summary and conclusions}
We have developed a nano-fabrication process for quantum photonic integrated circuits involving suspended and PhC waveguides without the deposition and etching of a hard mask. Particular attention has been devoted to achieving vertical holes in the ICP-RIE process and to clean all residues from dry and wet etching. A good accuracy in reproducing PhC patterns has been observed and it is confirmed by a statistical analysis of the hole-radii distribution.
The process described here could be extended to fabricate more complex devices involving electrical gates or on-chip superconducting detectors. In fact, the low-temperature process strongly minimizes the chances of damaging metallic contacts or sputtered films. Additionally, HCl or HF can be used almost interchangeably depending on the required material selectivity, so that dielectrics such as Si$_3$N$_4$, SiO$_2$ or flowable oxides could be used as well, e.g., to protect sensitive parts of the chip during dry or wet etching. 
Finally we should like to mention that our fabrication process is relevant for a range of applications of GaAs nanomembranes, e.g., quantum optomechanics \cite{ding_wavelength-sized_2011,usami_optical_2012}.

\ack
We gratefully acknowledge Financial support from the Carlsberg Foundation, the Lundbeck foundation, and the Danish Council for Independent Research (Natural Sciences and Technology and Production Sciences).
The authors would like to acknowledge H. El-Ella, C. B. S\o rensen, and S. Upadhyay for technical assistance.


\section*{References}
\begin{thebibliography}{10}

\bibitem{shields_semiconductor_2007}
Shields A J 2007 {\it Nat. Photonics} {\bf 1} 215--223
\bibitem{obrien_photonic_2009}
O'Brien J L, Furusawa A and Vu{\v c}kovi{\'c} J 2009 {\it Nat. Photonics.} {\bf 3} 687--695
\bibitem{rogers_synthesis_2011}
Rogers J A, Lagally M G and Nuzzo R G 2011 {\it Nature} {\bf 477} 45--53
\bibitem{benson_assembly_2011}
Benson O 2011 {\it Nature} {\bf 480} 193--199 
\bibitem{lodahl_interfacing_2013}
Lodahl P, Mahmoodian S and Stobbe S 2013 {\it arXiv:1312.1079 [cond-mat, physics:physics, physics:quant-ph]}
\bibitem{metcalf_quantum_2014}
Metcalf B J {\it et al.} 2014 {\it Nat. Photonics.} {\bf 8} 770--774
\bibitem{silverstone_-chip_2014}
Silverstone J W {\it et al.} 2014 {\it Nat. Photonics.} {\bf 8} 104--108 
\bibitem{hoang_enhanced_2012}
Hoang T B {\it et al.} 2012 {\it Appl. Phys. Lett.} {\bf 100} 061122
\bibitem{arcari_near-unity_2014}
Arcari M {\it et al.} 2014 \PRL {\bf 113} 093603  
\bibitem{fushman_controlled_2008}
Fushman I, Englund D, Faraon A, Stoltz N, Petroff P and Vu{\v c}kovi{\'c} J 2008 {\it Science} {\bf 320} 769--772
\bibitem{javadi_single-photon_2015}
Javadi A {\it et al.} 2015 {\it arXiv:1504.06895 [physics, physics:quant-ph]}
\bibitem{hennessy_quantum_2007}
Hennessy K {\it et al.} 2007 {\it Nature} {\bf 445} 896--899
\bibitem{loncar_design_2000}
Loncar M, Doll T, Vu{\v c}kovi{\'c} J and Scherer A 2000 {\it J. Lightwave Technol.} {\bf 18} 1402--1411
\bibitem{painter_two-dimensional_1999}
Painter O, Lee R K, Scherer A, Yariv A, O'Brien J D, Dapkus P D and Kim I 1999 {\it Science} {\bf 284} 1819--1821
\bibitem{gonzalez_fabrication_2014}
Gonz{\'a}lez I P {\it et al.} {\it J. Vac. Sci. Technol.} B {\bf 32} 011204
\bibitem{sprengers_waveguide_2011}
Sprengers J P {\it et al.} 2011 {\it Appl. Phys. Lett.} {\bf 99} 181110
\bibitem{hennessy_positioning_2004}
Hennessy K, Badolato A, Petroff P M and Hu E 2004 {\it Photonics and Nanostructures - Fundamentals and Applications} {\bf 2} 65--72
\bibitem{khankhoje_modelling_2010}
Khankhoje U K {\it et al.} 2010 {\it Nanotechnology} {\bf 21} 065202
\bibitem{atlasov_effect_2009}
Atlasov K I, Gallo P, Rudra A, Dwir B and Kapon E 2009 {\it J. Vac. Sci. Technol.} B {\bf 27} L21--L24
\bibitem{braive_inductively_2009}
Braive R {\it et al.} 2009 {\it J. Vac. Sci. Technol.} B {\bf 27} 1909--1914
\bibitem{baca_fabrication_2005}
Baca A G and Ashby C I H 2005 {\it Fabrication of GaAs devices} (The Institution of Electrical Engineers)
\bibitem{sun_selective_2009}
Sun X {\it et al.} 2009 {\it Solid-State Electronics} {\bf 53} 1032--1035
\bibitem{maboudian_critical_1997}
Maboudian R and Howe R T 1997 {\it J. Vac. Sci. Technol.} B {\bf 15} 1--20
\bibitem{voncken_etching_2004}
Voncken M M A J {\it et al.} 2004 {\it J. Electrochem. Soc.} {\bf 151} G347--G352
\bibitem{desalvo_wet_1996}
DeSalvo G C {\it et al.} 1996 {\it J. Electrochem. Soc.} {\bf 143} 3652--3656
\bibitem{ding_wavelength-sized_2011}
Ding L, Baker C, Senellart P, Lemaitre A, Ducci S, Leo G and Favero I 2011 {\it Appl. Phys. Lett.} {\bf 98} 113108
\bibitem{usami_optical_2012}
Usami K, Naesby A, Bagci T, Melholt Nielsen B, Liu J, Stobbe S, Lodahl P and Polzik E S 2011 {\it Nat. Phys.} {\bf 8} 168--172
\end{thebibliography}
\end{document}